\documentclass[]{aastex701}

\begin{document}

\title{JWST Characterization of Earth Quasi-Satellite (469219) Kamo`oalewa}

\author[0000-0003-1383-1578]{Benjamin N. L. Sharkey}
\affiliation{Department of Astronomy, University of Maryland \\
College Park, MD 20742-2421, USA}
\email[show]{sharkey@umd.edu}

\author[0000-0002-7743-3491]{Vishnu~Reddy}
\affiliation{Lunar and Planetary Laboratory, University of Arizona, Tucson, AZ, USA}
\email{vishnureddy@arizona.edu}

\author[0000-0003-2152-6987]{John W. Noonan}
\affiliation{Department of Physics, Edmund C. Leach Science Center, Auburn University, Auburn, AL, USA}
\email{noonan@auburn.edu}

\author[orcid=0000-0003-1008-7499, sname='T. Kareta']{Theodore Kareta}
\affiliation{Department of Astrophysics and Planetary Science, Villanova University, Villanova, PA, USA}
\email{theodore.kareta@villanova.edu}  

\author[0000-0001-9542-0953]{James M. Bauer}
\affiliation{Department of Astronomy, University of Maryland \\
College Park, MD 20742-2421, USA}
\email{gerbsb@umd.edu}

\author[0000-0002-6117-0164]{Bryan J. Holler}
\affiliation{Space Telescope Science Institute, Baltimore, MD, USA}
\email{bholler@stsci.edu}

\author[0000-0002-9042-408X]{Yaeji Kim}
\affiliation{Department of Astronomy, University of Maryland \\
College Park, MD 20742-2421, USA}
\email{ykim1231@umd.edu} 

\author[0000-0003-2872-0061]{Albert R. Conrad}
\affiliation{Large Binocular Telescope Observatory, Tucson, AZ, USA} \email{aconrad@lbto.org}

\begin{abstract}

Near-Earth asteroid (469219) Kamo`oalewa is a uniquely stable quasi-satellite of the Earth and a target of the Tianwen-2 spacecraft mission. Here we report observations taken with JWST's NIRSpec instrument in integral field unit (IFU) mode in February 2026. The JWST reflectance spectrum is notably less red (more neutral) from $1.0-2.5$ $\mu m$ than previous ground-based spectrophotometric observations. New observations made with LBT in April 2026, observed and processed similarly to the 2021 observations, find $zJ$ colors in agreement with JWST. Kamo`oalewa's infrared colors appear more similar to typical silicate-rich asteroids and unlike the reddened, space-weathered lunar-like silicates suggested by previous observations. In agreement with the ground-based spectrum, we detect a faint silicate absorption feature at $0.93\pm 0.01$ $\mu m$. We do not detect a 2.0 $\mu m$ silicate absorption. Models of Kamo`oalewa's faint thermal emission (beginning near 4.5 $\mu m$) find a mean diameter of $D=18\pm2\mathrm{m}$ and best-fit visible albedo $p_V = 0.59^{+0.25}_{-0.17}$, with models as low as $p_V = 0.36$ providing adequate model fits. This combination of color, albedo, and absorption bands is similar to oldhamite-bearing enstatite-rich compositions. Kamo`oalewa's brightness variations over the course of the JWST program provides independent confirmation of its rotation period of 27.9 minutes, with an axis ratio $\sim1.4$ ($D\sim15-21$ m). 

\end{abstract}



\section{The Origins of Earth Quasi-Satellites}

Earth quasi-satellites are a subcategory of co-orbital objects that remain continuously close to the Earth on heliocentric orbits. Due to their close geocentric distances and low delta-V, quasi-satellites are of high interest for spacecraft exploration and possible resource utilization. (469219) Kamo`oalewa is the most stable known example, persisting in its current quasi-satellite configuration for hundreds of years \citep{Marcos2016}. Other Earth co-orbitals are generally stable less than 10 years and have unfavorable geometry for repeated telescopic observations. Kamo`oalewa's long-term behavior is poorly constrained at present, but it is likely to remain in a co-orbital state for at least 0.5 Myr, and possibly for several Myr depending on its size and nongravitational forces like the Yarkovsky effect \citep{Fenucci2021}. Kamo`oalewa is a target for the forthcoming Tianwen-2 sample return mission \citep{Hu2023,Jin2019,Li2019}, which will test Kamo`oalewa's origins.

Earth quasi-satellites were initially presumed to simply be small main-belt asteroids captured by chance into very near-Earth orbits, similar to the baseline near-Earth asteroid (NEA) population. Initial visible colors \citep{Reddy2017} supported this view. Subsequent ground-based characterization \citep{Sharkey2021} found that Kamo`oalewa's visible spectrum matches a variety of silicate-rich asteroid types, but found a distinctly red spectral slope in the infrared ($zJHK$) that differed from almost any NEA. Kamo`oalewa's infrared colors were found to more closely match space-weathered silicates, similar to those present in the lunar regolith. Dynamical models conclude certain rare ejecta scenarios can reproduce Kamo`oalewa's present orbit \citep{CastroCisneros2023}, and impact models suggest Giordano Bruno crater could have ejected Kamo`oalewa-sized fragments \citep{Jiao2024}. However, dynamical models of main-belt implantation also satisfactorily (and more frequently) produce Kamo`oalewa-like objects \citep{Fenucci2026}, illustrating the competing pathways for Kamo`oalewa's origins.

JWST observations of Kamo`oalewa were conducted in February 2026 using the NIRSpec instrument in IFU prism mode ($0.7-5.1$ $\mu m$). These observations were collected to provide a pre-encounter reflectance spectrum and a sparse lightcurve. The NIRSpec observations cover a range of key reflectance features between $0.7-4.0$ $\mu m$ without confounding effects from telluric absorption, enabling key compositional tests and comparison to previous observations. The NIRSpec wavelength range also covers the start of thermal emission beyond $4\mu m$, enabling the first constraints on Kamo`oalewa's size and albedo.

\section{JWST Data Processing}

Raw {$\it uncal$} data were downloaded from the Mikulski Archive for Space Telescopes (MAST) and locally processed with version 1.20.2 of the JWST calibration pipeline ({$\tt jwst$}; \citealt{Bushouse2025}) and reference file context {$\it jwst\_1477.pmap$}. The fully calibrated {$\it s3d$} data cubes were then reduced following the template PSF fitting method first described in \citet{Wong2024} and in specific detail in \citet{Sharkey2025}. Compared to the process followed in \citet{Sharkey2025}, an additional step was performed to remove lingering instrumental artifacts from the extracted 1D spectra. The Kamo'oalewa spectra were first divided by the observed spectrum of the G3V standard star SNAP-2 \citep{Gordon2022}, then multiplied by the CALSPEC model of SNAP-2 \citep{Bohlin2014} convolved to the resolving power of the NIRSpec Prism. This step removed any remaining artifacts while recovering the original spectral shape. Then, the solar component was removed by dividing by a solar model retrieved from the Planetary Spectrum Generator (PSG; \citealt{Villanueva2018}) and convolved to the resolving power of the NIRSpec Prism. The combination of dithers and removal of outliers then proceeded as described in \citet{Sharkey2025}.

Kamo`oalewa was observed over seven visits from Feb. 9--15, 2026. Each visit was divided into four dithers, with an effective exposure time of 641.911 seconds per dither. The target fell outside of the NIRSpec IFU field-of-view for one dither during the Feb. 14 observations, with two other dithers with the target close to the edge. For one additional visit (visit ``OBS5''), Kamo`oalewa was outside of the field-of-view for all four dithers. The full time span of each visit (the start of the first dither to the end of the last dither) was approximately 52 minutes (51:47:506), the maximum duration before triggering a new visit due to guide star constraints.

Spectral analysis is based on median-combining each dither into a single ``visit spectrum," with the uncertainty on a wavelength bin in the visit spectrum taken as the standard deviation of the four dither measurements. These visit spectra were further combined into a grand-average spectrum, taken as the median combination of each visit, with errors computed as the error-on-the-median. To produce a lightcurve, reflected brightnesses were computed on a per-dither basis, totaling 27 observations of 641.911 seconds each.

\section{Lightcurve}
The 28 dithers were analyzed for variations in Kamo`oalewa's reflected brightness. Reflected fluxes were determined by computing medians and uncertainties across binned wavelength elements. Various bin widths were tested, and 75-element (width 0.375 $\mu m$) bins were ultimately chosen to provide an adequate sample of points for computing the standard deviation given the achieved per-dither SNR. Median values of bins between $0.7-3.5$ $\mu m$ were then summed to produce the values shown in Table \ref{circumstances}. Typical uncertainties range from $3-6$\%.

Integrated brightnesses and exposure mid-times from Table \ref{circumstances} were scaled to relative magnitudes and analyzed via Lomb-Scargle periodogram implemented in {\tt astropy} ({$\tt astropy.timeseries.LombScargle$}). We adopted a two-term sinusoidal fit using the ``chi2'' method, with a period search from $25-150$ cycles/day and 100 samples/peak. Uncertainties were computed by bootstrap resampling the magnitude values according to their Gaussian uncertainties, recomputing the periodogram, and returning the frequency with peak power. The bootstrap process was iterated 1000 times. 

\begin{figure*}[ht!]
\plotone{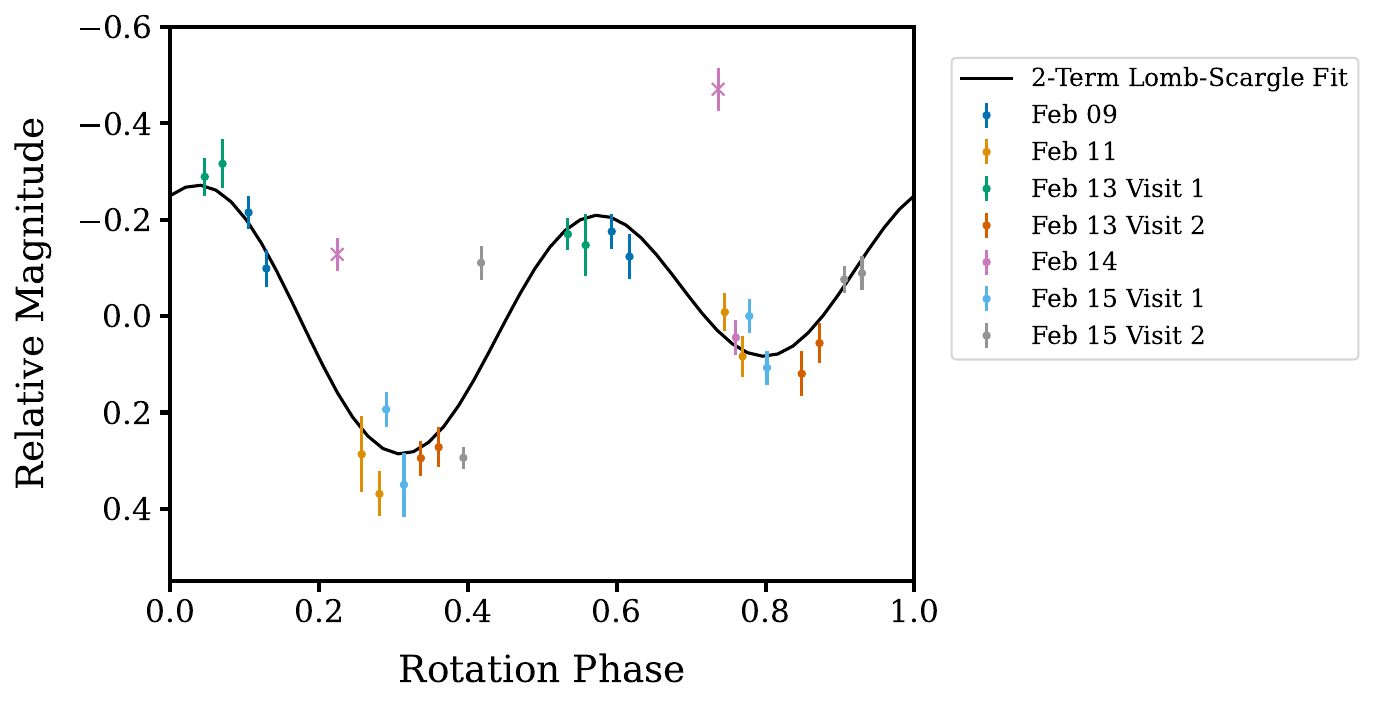}
\caption{Top: Kamo`oalewa's lightcurve, phased to best-fit $P = 27.900^{+0.002}_{-0.003}$ minutes. The JWST observations display asymmetric minima, consistent with prior ground-based results. Outliers excluded from period fits are marked with 'X' symbols.
\label{fig:lightcurve}}
\end{figure*}

Using only the JWST data, we find a period of $27.900_{-0.003}^{+0.002}$ minutes (16-84th percentile ranges), producing a double-peaked asymmetric lightcurve. The phase-folded lightcurve is displayed with the best-fit sinusoid in Figure \ref{fig:lightcurve}. The two Feb. 14 dithers where Kamo`oalewa is close to the edge of the field-of-view are clearly discordant with the remaining data, and we exclude them from the period fit. These outlier points are identified and displayed in Figure \ref{fig:lightcurve}. 

We find that, despite the random and sparse cadence of observations (per telescope scheduling constraints), and the long integration times of each dither relative to Kamo`oalewa's rotational variation, JWST's photometric stability provides a period estimate robust against aliasing. This is due to Kamo`oalewa's large lightcurve amplitude, easily resolved within the SNR, and our sampling of the asymmetric brightness minima and maxima across multiple visits. Previous observations \citep{Sharkey2021} noted a significant alias at $P\sim14$ minutes, but disfavored this solution due to Kamo`oalea's observed lightcurve asymmetry. 

Our bootstrapped period search favors the 27.9 minute period: of the 1000 runs, $\sim88\%$ of solutions cluster around 28 minutes, while $\sim12\%$ cluster around 14 minutes. Our preferred best-fit period of $27.900_{-0.003}^{+0.002}$ minutes is consistent with previous observers, from Kamo`oalewa's discovery \citep[$P<30$min][]{Tholen2016}, to initial characterization \citep[$P=28.3^{+1.8}_{-1.3}$ min][]{Sharkey2021} and recent reanalysis of data from 2016-2018 \citep[$P=27.9$min][]{Bonamico2026}.

The JWST lightcurve amplitude of $0.7 \pm 0.1$ magnitudes is smaller than the value of 1.07$\pm$0.05 magnitudes from \citet{Sharkey2021}. The different lightcurve amplitudes suggest that the two observations sampled distinct surface profiles across different sub-observer latitudes. Further knowledge of Kamo`oalewa's pole orientation and shape will determine the regional overlap between the ground- and space-based vantages.

\section{Comparisons to Ground Reflectance Data}

The NIRSpec data clearly diverge from the composite ground-based spectrum collected by \citet{Sharkey2021}. To quantify these differences and investigate possible causes, we focus our comparisons on the spectral region from $0.95-1.65$ $\mu m$. Specifically, we compare the NIRSpec data to the colors derived at $z$, $J$, and $H$ bands.

The 2021 photometric observations from \citet{Sharkey2021} were collected as paired simultaneous images in two bands ($z$, $J$, $H$, or $K$). This methodology controlled for lightcurve variations between different filters, eliminating this as a possible systematic source of uncertainty. Object fluxes were stacked at multiple dither locations, with background subtraction applied using frames within five minutes of each exposure. Stacks of frames at each dither position were produced, and these stacks were compared with one another to ensure that each dither position provided similar (within uncertainty) flux measurements. This comparison procedure provides resilience to contamination from faint background objects. Due to Kamo`oalewa's faintness, observations were stacked over $30-60$ minute intervals, with the instrumental fluxes in each band converted to relative reflectance via subsequent measurements of solar analog star GSPC P330-E.

\begin{figure*}[ht!]
\centering

\includegraphics[width=0.45\textwidth]{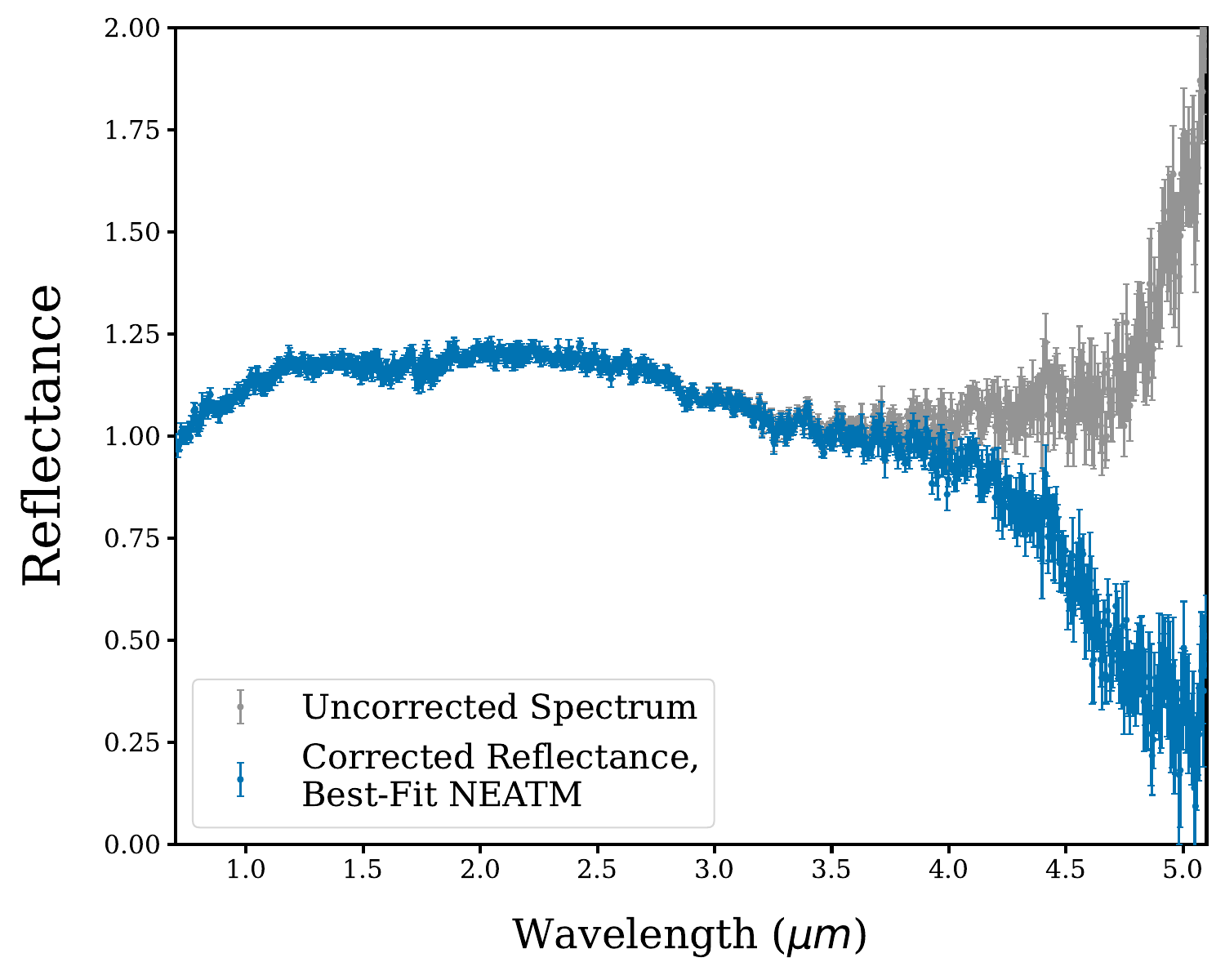}

\includegraphics[width=0.45\textwidth]{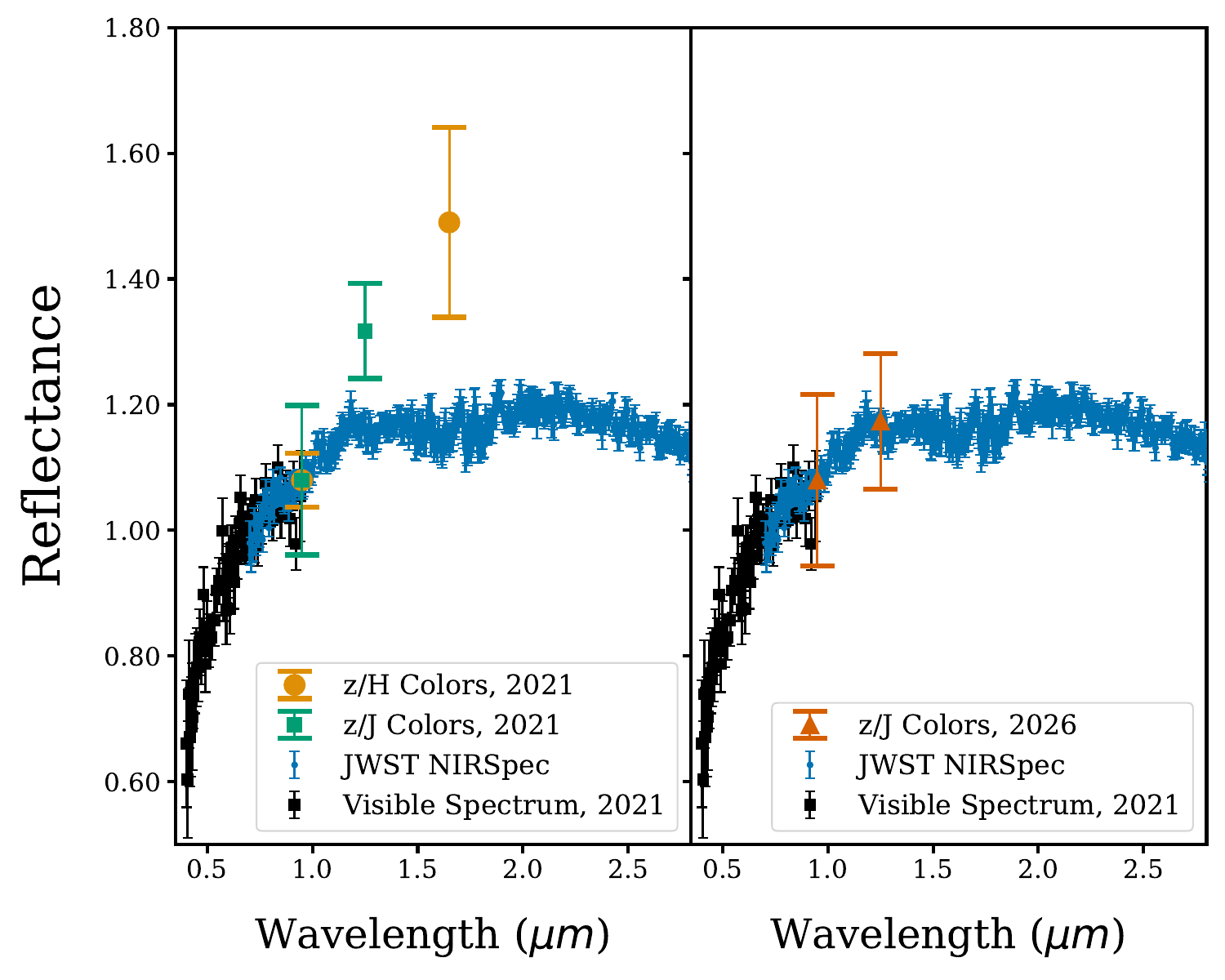}

\includegraphics[width=0.45\textwidth]{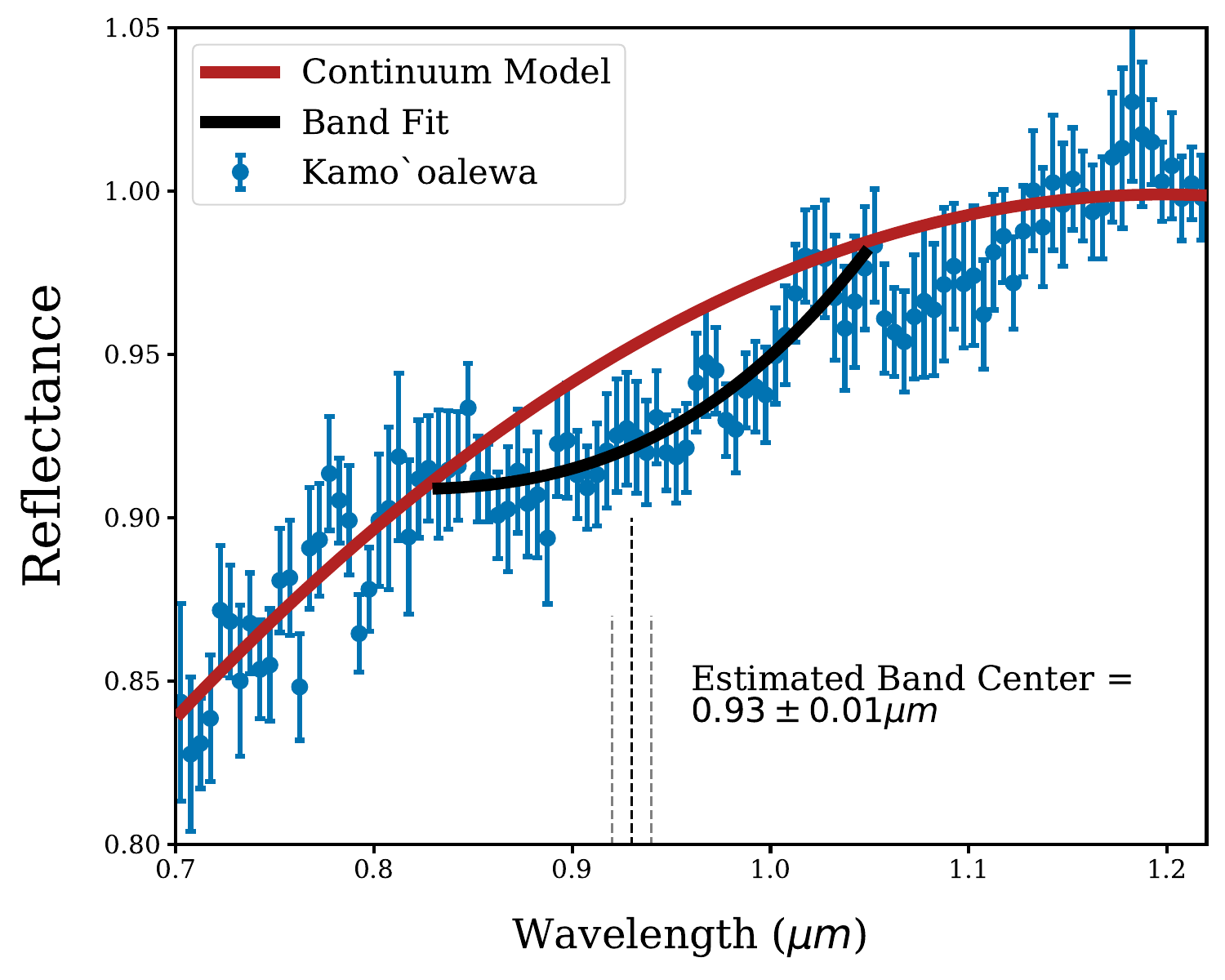}
\caption{Top: NIRSpec reflectance spectrum, including fluxes corrected for thermal emission (blue) and the original uncorrected spectrum (gray). The thermal correction was taken by subtracting the best-fit NEATM model from the grand-averaged SED. The overall spectrum is mildly red from 0.7-2.5$\mu m$, and is blue at wavelengths greater than 2.5$\mu m$. Middle: Comparison of JWST and LBT near-infrared measurements. The 2021 observations have higher reflectance values at $J$ and $H$ than the JWST spectrum, demonstrating a clear discrepancy between JWST and previous ground-based infrared colors. Follow-up $zJ$ measurements in 2026, conducted with the same experimental setup as in 2021, finds good agreement with JWST but with limited SNR due to observing conditions. Bottom: We identify an absorption feature near 0.9$\mu m$ which we associate with pyroxenes. The continuum model is displayed as a red polynomial fit to the spectrum outside of the band. The band fit is plotted in black, with resulting band center of $0.93\pm0.01\mu m$.
\label{fig:spectra}}
\end{figure*}

Figure \ref{fig:spectra} displays the $zH$ and $zJ$ measurements from the 2021 observations of \citet{Sharkey2021}. The $zJ$ frames from 2021 were not originally processed and analyzed, as they were a short observation originally obtained to locate the object in the background field for subsequent images at $H$ and $K$ bands. Both the $zJ$ and $zH$ color measurements from 2021 suggest a steeply reddening slope from $1.0-1.65$ $\mu m$, though with large uncertainties ($> 10\%$).

In April 2026, additional measurements were conducted using LBT single-telescope observations with the LUCI2 instrument. The LBT observations included a validation with the bright asteroid Eulalia, recently observed with JWST as part of program GO 6384 (PI: Takir). Using the same data reduction methods, and the same solar analog star, we derive $zJH$ colors for Eulalia that are consistent within $3 \%$ of JWST observations of Eulalia \citep{McClure2026}. This gives us confidence in the overall experimental design and data reduction protocols. The re-observations of Kamo`oalewa with LBT have limited SNR due to weather conditions (which also precluded $H$ band observations), but the derived $zJ$ colors agree with the JWST spectrum. However, we note that the $zJ$ measurements from 2021 and 2026 are also consistent with each other, highlighting the current SNR limits of tests between data sets. Additional investigation of possible systematic and physical circumstances, particularly background estimations, between the 2021 and JWST colors are clearly warranted, as Kamo`oalewa's infrared color is a major compositional constraint that could help confirm various origin scenarios. The cause of the color discrepancy remains an open question.

\section{Size Determination and Thermal Modeling}

\subsection{Modeling Description}\label{ModelDesc}

To determine Kamo`oalewa's diameter and visible albedo, we employ a combined model of its reflected light near $0.8\ \mu m$ and its thermally emitted light near $5.0\ \mu m$. To fit the thermal flux, we employ the NEATM formalism \citep{Harris1998}. NEATM predicts thermal fluxes according to the Sun-target-observer phase angle $\alpha$, the object's size, its beaming parameter, $\eta$, and its bolometric Bond albedo, $A$. These parameters define the target's surface temperature as a function of latitude and longitude, $T(\phi, \theta)$, and the flux received by the observer, $F_{\lambda}$ \citep{Harris1998,Delbo2002}: 

\begin{equation}
F_{\lambda} = \frac{\epsilon D^2}{\Delta^2} \frac{hc^2}{\lambda^5} \int\limits_{0}^{\pi/2} \int\limits_{-\pi/2}^{\pi/2} \frac{cos^2(\phi)cos(\theta-\alpha))}{exp(\frac{hc}{\lambda k T(\theta,\phi)})-1} \,d\theta\,d\phi
\end{equation}

with zero emission on the nightside, where $cos(\theta-\alpha)<0$. The temperature distribution of the surface is described by

\begin{equation}
T(\theta,\phi)= \left[\frac{(1-A)S_\odot}{\epsilon\sigma\eta}\right]^{1/4}(cos\phi)^{1/4}(cos\theta)^{1/4}.
\end{equation}

Per standard definition of the phase function $\Phi$, the reflected flux of an object at a given wavelength $\lambda$ at heliocentric distance $r$ with geometric albedo $p_V$ observed at distance $\Delta$ \citep[see][equation A.8 for a recent derivation]{MacLennan2026} is

\begin{equation}
    F_{refl,\lambda} = p_V \frac{F_{\odot,\lambda}}{r^2}\left(\frac{D/2}{\Delta}\right)^2\Phi(\alpha),
\end{equation}

\noindent where $F_{\odot,\lambda}$ is the solar irradiance. We take the model solar spectrum ({\it sun\_reference\_stis\_002.fits}) from the CALSPEC database \citep{Bohlin2014} , and adopt the $H$,$G$ magnitude system of \citet{Bowell1989}.

The NEATM models have several free parameters: object diameter $D$, Bond albedo $A$, beaming parameter $\eta$, and the total thermal contribution (see Section \ref{ThermalContribution}). Models with combined thermal and reflected light fits add two free parameters: the phase-slope parameter $G$, and the geometric albedo ratio $p_{0.8\mu m}/p_V$. Such thermal models are clearly highly degenerate, particularly in this case with weak thermal flux overlaid on a rapidly-changing reflectance curve. Our goal is not to return a single best-fit model, but instead to bracket the range of possible diameters and albedos for various model assumptions. We employ MCMC models with flat priors for $1\ m<D<100\ m$, $0.01<A<1.00$, and $0<\eta<\pi$.

Thermal models are typically constrained by knowledge of the absolute magnitude, $H$. However, Kamo`oalewa's known large lightcurve amplitude complicates comparisons to its reported value of 24.3, which does not specify the rotation phase. Instead, we use the reflected-light portion of the NIRSpec spectral energy distribution (SED) from $0.75-0.85$ $\mu m$. Estimating the phase slope $G$ connects NEATM-derived $D$ and $A$ to $p_V$.  The phase slope parameter $G$ is often assumed to be 0.15 for asteroids with unmeasured phase curves, but this assumption is not valid for high albedo asteroids \citep[mean $G\sim0.5$,][]{Harris1989}, nor necessarily for 10-m class objects: recent observations of 2024 YR$_4$, a similarly small and rapidly rotating NEA recently studied with JWST has $G=0.51 \pm 0.11$ \citep{Devogele2026}. We therefore allow $G$ to be fit with a flat prior of $-0.2<G<1.0$ to encompass a wide range of low- and high-albedo cases \citep{Li2015,Belskaya2000,Hicks2014}. 

We estimate the geometric albedo ratio $p_{0.8\mu m}/p_V=1.16 \pm 0.02$ from the publicly available visible spectrum of \citet{Sharkey2021}, and allow models to explore a wider range of values using a Gaussian prior of $p_{0.8\mu m}/p_V =  1.16 \pm 0.06$. All MCMC models employed 32 walkers for 5000 iterations, with the first 1000 iterations discarded as burn-in steps. All quoted best fits and uncertainty ranges in posterior parameters represent the median value and the $16-84$ percentile range. 

\subsection{Limits on Thermal Contributions from 4.5-5.2 $\mu m$}\label{ThermalContribution}

Kamo`oalewa's thermal emission must be isolated from its reflected flux in order to return proper parameter retrievals with NEATM. Thermal emission is commonly estimated by extrapolating the SED from wavelengths dominated by reflected light. Simply treating Kamo`oalewa's reflectance as flat (solar) or as having a constant linear slope from $4.0-5.1$ $\mu m$ produces an unreasonably faint emission spectrum that can only be fit by physically implausible models for rocky materials (i.e., geometric albedos $>$ 0.9, $\eta=\pi$). 

\begin{figure*}[ht!]
\plotone{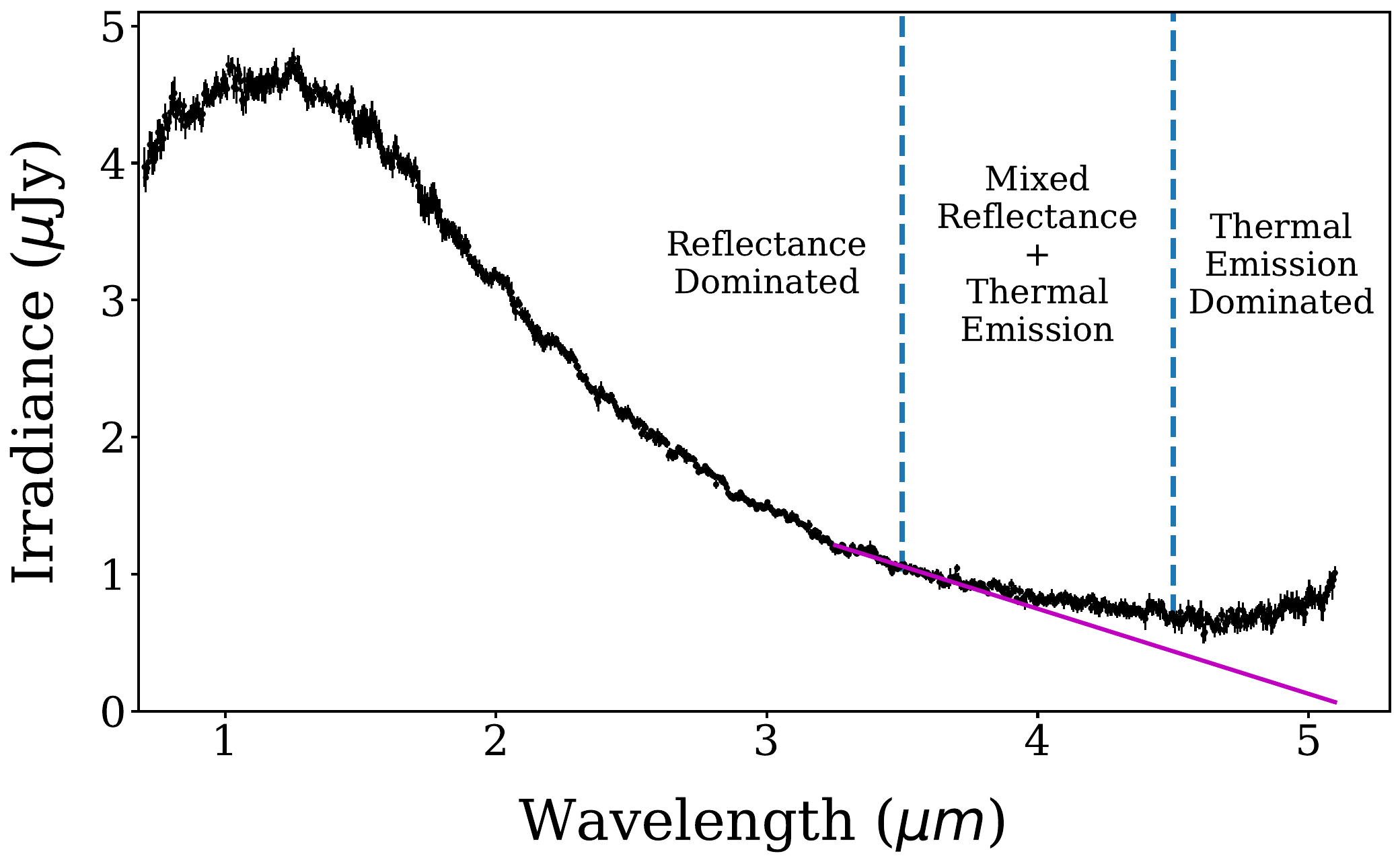}
\caption{Kamo`oalewa's irradiance averaged across all JWST observations. Approximate regions for reflectance and thermal emission are noted. Kamo`oalewa displays very little thermal emission short of 4.5 $\mu m$. We linearly extrapolate the reflected flux near 3.25-3.75$\mu m$ in order to constrain the contributions from emitted fluxes at $\lambda>4.5\mu m$. The slope of this extrapolation is fit by our models, with a typical example illustrated in magenta.
\label{fig:irrad}}
\end{figure*}

However, linear extrapolation of the irradiance spectrum ($F_\lambda$, not the reflectance spectrum) from $\sim 3.25-3.75$ $\mu m$ does produce plausible emission curves (see magenta curve of Figure \ref{fig:irrad} ). This thermal flux estimate implies that Kamo`oalewa's reflectance changes nonlinearly from $4-5$ $\mu m$. We adopt this extrapolation technique for our thermal model fits, but note that it is a major assumption imposed on the data. Our fits allow adjustment of the linear extrapolation to allow the thermal flux at 5.1 $\mu m$ to vary from $0-100$$\%$ of the measured total. Across the range of model assumptions explored in this work, fits consistently find the contribution of thermal vs reflected flux at 5.1 $\mu m$ to be $\sim 80-100\%$. We therefore estimate an additional per-point 0.0002 mJy systematic uncertainty ($\sim 25 \%$ of the peak flux at 5.1 $\mu m$) which is included in all thermal fits. To limit contamination from improperly removed reflected light, we constrain models to fit fluxes from $4.5-5.1$ $\mu m$ only.

To evaluate our estimation of thermal emission, we can consider the case where reflected irradiance decreases faster than a linear extrapolation beyond 4 $\mu m$. Such a case would result in increased emission at shorter wavelengths resulting in lower values of Bond albedo. The most extreme version of this case is one where emission entirely dominates beyond 4.5$\mu m$. Models fit to Kamo`oalewa's total flux from 4.8-5.1$\mu m$ (without any subtraction of reflected flux) return solutions of $D=17.3^{+3.3}_{-3.2}\ m$ with $p_v>0.3$ (first percentile). 

The opposing limit of less thermal emission is accommodated in our extrapolation fitting. Assuming that Kamo`oalewa emits less thermal flux than is preferred by our models leads to lower diameters and higher albedos. This limit of less thermal emission is better constrained by Kamo`oalewa's reflected flux. We illustrate this effect by considering underconstrained models without the reflected flux constraint. Assuming a $G$ value of 0.5, and requiring $A$ such that $0<p_V<1.0$, we return valid solutions for  $D=11^{+13}_{-6}$ m and $p_V = 0.27^{+0.49}_{-0.23}$. But such models include clearly unrealistic solutions. The absolute magnitude for an object of $D$ = 11 m and $p_V=1$ is $H$ = 25.4, one magnitude fainter than Kamo`oalewa's reported value. By incorporating Kamo`oalewa's JWST SED from $0.75-0.85$ $\mu m$, we can exclude such implausible solutions.

\subsection{Constraining Diameter and Albedo}

\begin{figure*}[ht!]
\plotone{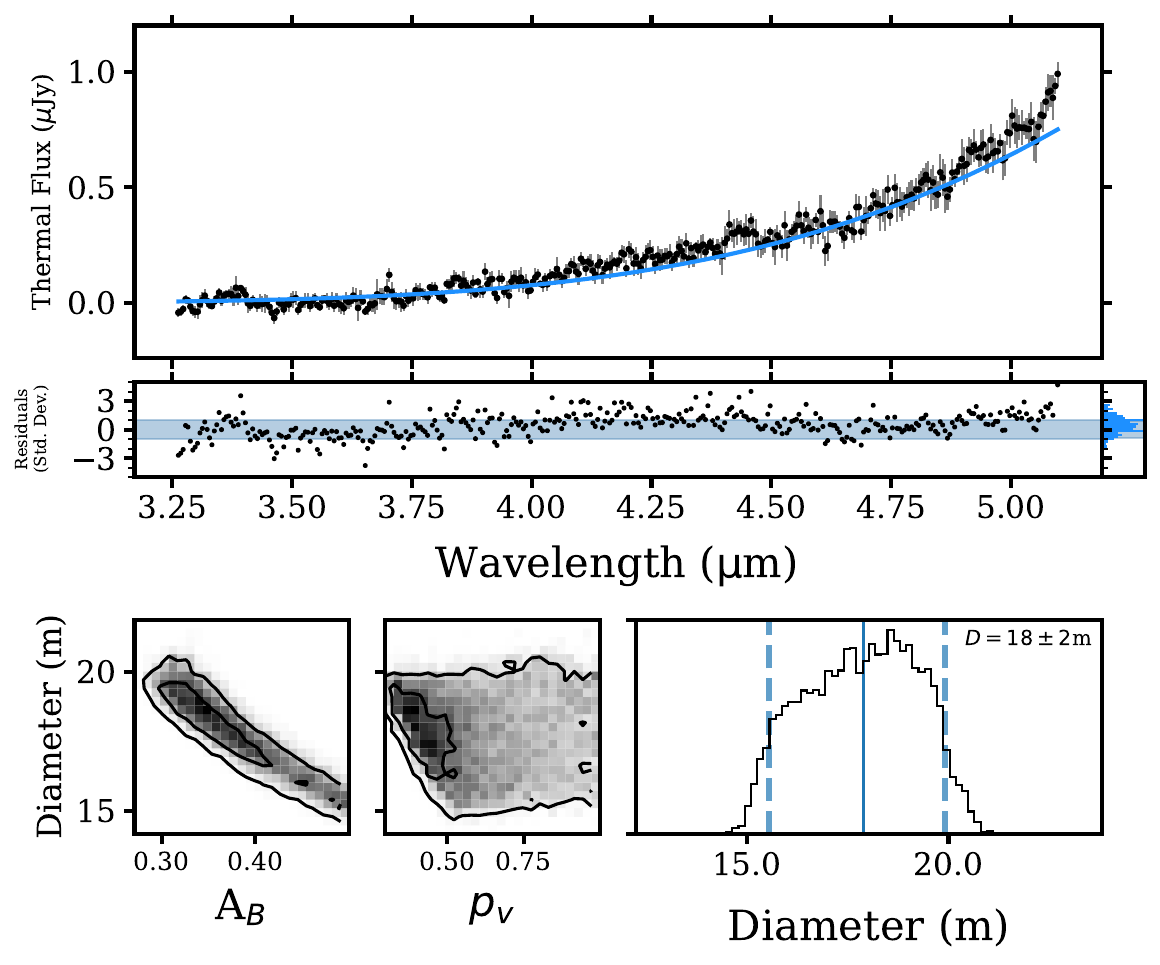}
\caption{NEATM model fits, including distributions of Bond albedo $A$, visible geometric albedo $p_V$, and diameter $D$. Model fits were performed on the averaged irradiance spectrum after scaling each visit to the overall flux of the final set of observations (Feb. 15).
\label{fig:NEATM_fits}}
\end{figure*}

Unlike larger, slower rotating NEAs, Kamo`oalewa's observed SED contains very little thermal emission at $\lambda <$ 4 $\mu m$ relative to its overall brightness. Our fitting technique, including fitted extrapolation of thermal contribution and fits to reflected fluxes at 0.8 $\mu m$, returns our preferred solution. This preferred model retrieves $D$ = 18$\pm$2 and $p_V = 0.59^{+0.25}_{-0.17}$, with $\eta$ varying from our imposed upper limit of $\pi$ to $\eta = 2.19$ (16th percentile). The preferred model curve, residuals, and diameter/albedo distributions are displayed in Figure \ref{fig:NEATM_fits}. Diameter, $A$, and $\eta$ are highly correlated, with the lower limit of diameter corresponding to the lower limit of $\eta$ and the upper limit of $A$. The lower limit of $A$, where $\eta\sim\pi$, corresponds to $p_V\sim0.36$.

The case where $\eta=\pi$ places the subsolar temperature equal to that of the fast-rotator model \citep[FRM,][]{Lebofsky1989}. Finding good fits using such extreme cases of $\eta$ illustrates how little emission we measure from Kamo`oalewa's surface. Kamo`oalewa's fast rotation period supports such high values of $\eta$, though  Kamo`oalewa's unknown obliquity could diverge from FRM assumptions. While the temperature distributions of NEATM and FRM differ, Kamo`oalewa's observed phase ($\sim 60^{\circ}$) coincides with conditions found by \citet{Mommert2018} where both FRM and NEATM retrieve similar physical parameters. Indeed, validations using the FRM implementation of \citet{Delbo2002} finds best fits with similar physical parameters ($D=21\pm6\ \mathrm{m};\ p_V=0.6\pm0.3$), with uncertainties driven by $G$ and $H$. 

\section{Constraining Mineralogy}

Silicate-bearing asteroids often display absorption features near 1.0 and 2.0 $\mu m$, dubbed ``Band I'' and ``Band II.'' Using the methodology of \citet{Sanchez2020} (adapted from \citet{Kareta2025}), we detect and characterize a faint feature at 0.93 $\pm 0.01$ $\mu m$ with a depth of $3.0 \pm 0.5\%$. There are several other weak (1-2$\%$) features in the JWST spectrum (for example, near 3.0 $\mu m$), but their significance depends on choice of continuum definition in regions with lower SNR compared to the 1.0 $\mu m$ region. Future investigation of these features will require assessing whether they are robust across each visit.

The measured Band I center and modeled albedo constrains the plausible surface mineralogy of Kamo`oalewa. Previous assessment of Kamo`oalewa's mineralogy \citep{Sharkey2021} interpreted the quasi-satellite's steep infrared colors to favor compositions akin to dark, lunar-like silicates \citep[$p_v\lesssim0.1,$ e.g.][]{Pieters2000,Taylor2001,Lin2020}) instead of moderately brighter S-type asteroids like (16) Ausonia \citep[$p_v\sim0.16$][]{Davis2005}. Previous comparisons used the approximate albedo range between reddened lunar silicates and Ausonia (0.10-0.16), adopting a nominal H=24.3, to return a size of 46-58m. However, these previous size estimates were based only on compositional comparisons without independent albedo  constraints. Our modeled albedo range presents new compositional interpretations. 

Our modeled albedo range in this work allows for three possible surface mineralogies. These are moderate albedo S- and V-types and high albedo E-types. S-types have typical albedos between 0.15-0.30, V-types between $0.3-0.45$ and E types have albedos between $0.40-0.70$ \citep{Cantillo2026,Sanchez2025}. Surfaces of typical S-type asteroids that have surface mineralogies similar to ordinary chondrite meteorites are dominated by the minerals olivine and pyroxene \citep{Reddy2015}. A mixture of these minerals produce distinctive absorption bands at $\sim1.0$ $\mu m$ and $\sim2.0$ $\mu m$ that have band depths between $\sim10-30\%$. Similarly, HED meteorites are the most plausible meteorite analogs for V-type asteroids. Reflectance spectra of HEDs are dominated by pyroxene absorption features at $\sim0.90$ and $2.0\ \mu m$ \citep{Reddy2012}. While Kamo`oalewa's modeled albedo range overlaps with S and V types, the lack of a 2-$\mu m$ absorption feature in Kamo1oalewa's reflectance spectrum suggests that its surface mineralogy does not match a typical S- or V-type. 

The albedo range for E-type asteroids \citep[$p_V>0.3$][]{Tholen1989}, whose surface mineralogy is similar to enstatite achondrites or aubrites, overlaps with the modeled albedo of Kamo`oalewa. Mineralogy of aubrites is dominated by enstatite which is a low-iron pyroxene that typically has a single weak ($ < 10\%$) absorption feature at $\sim0.90\ \mu m$ and no 2.0-$\mu m$ feature. In addition, aubrites have red spectral slope (increasing reflectance with increasing wavelength) between 0.35 to 0.70 $\mu m$ and blue spectral slope between 0.7 and 2.5 $\mu m$ \citep{Cantillo2024}. While the lack of the 2-$\mu m$ feature, high albedo, and spectral slopes match what is observed on Kamo‘oalewa, the measured Band I center for the asteroid at $0.93 \pm 0.01$ $\mu m$ is longer than Band I centers measured in laboratory spectra of aubrites, which typically have Band I centers between $0.89-0.91$ $\mu m$ \citep{Cantillo2026} due to low-Fe pyroxene (enstatite). However, in addition to low-Fe pyroxene, aubrites also have an opaque sulfide mineral called oldhamite (CaS). Oldhamite has a weak absorption feature at $\sim0.96$ $\mu m$ \citep{Burbine2002} that could move the Band I center of a typical low-Fe pyroxene absorption feature to longer wavelengths as seen on Kamo`oalewa. In fact, \citet{Gaffey2004} note that low Fe content in enstatite paired with a higher oldhamite abundance may shift the Band I center to longer wavelengths closer to the 0.96 $\mu m$ center. Based on these factors, it is plausible that the surface of Kamo`oalewa is similar to E-type asteroids.

\section{Constraining Kamo`oalewa's Internal Structure}
Given the new thermal and lightcurve-derived size information, we estimated the minimum cohesive strength required for Kamo`oalewa to maintain its present shape against rotational and self-gravitational stresses. We take a homogeneous prolate spheroid with dimensions of approximately 24$\times$17$\times$17 m, corresponding to an equivalent diameter of $\sim$20 m and an axis ratio of $\sim$1.4, rotating with the JWST-derived period of 27.900 min. Assuming a bulk density of 2000 kg m$^{-3}$, representative of silicate-rich asteroids \citep[e.g.,][]{fujiwara2006rubble}, we evaluated the internal stress field and the corresponding Drucker–Prager yield criterion. The maximum required cohesion was found to be only $\sim$0.24 Pa.

This value is extremely small compared to cohesive strengths commonly invoked for rubble-pile asteroids \citep[e.g.,][]{hirabayashi2015stress,scheeres2018implications}. Models of regolith-mediated cohesion predict bulk strengths ranging from a few pascals to hundreds of pascals depending on the characteristic grain size distribution, while competent rock exhibits strengths many orders of magnitude larger. The inferred cohesion requirement for Kamo`oalewa is therefore effectively negligible. 

In practical terms, the observed shape and spin state can be sustained by either a monolithic object or an extremely weak gravitational aggregate. Consequently, the current rotational state does not provide a meaningful constraint on Kamo`oalewa’s origin or internal structure. Additional information from Tianwen-2, including direct measurements of surface properties, internal structure, and material strength, will be required to determine whether Kamo`oalewa is a coherent fragment or a weakly bound aggregate.

\section{Conclusions}
We demonstrate JWST's power to simultaneously characterize the rotation state, near-IR reflectance, and thermal emission of Kamo`oalewa. Similar efforts typically require several distinct observing modes, often with different telescopes, over the course of multiple apparitions \citep[as in][]{Sharkey2021}. Using prism IFU mode with 28 sparse photometric points, we successfully recover the rotation period recovered by previous observers \citep{Tholen2016,Sharkey2021,Bonamico2026}, and find a smaller lightcurve amplitude ($0.7 \pm 0.1$ mag) than recorded from ground-based viewing geometries.  

The NIRSpec-derived SED for Kamo`oalewa favors NEATM solutions with $\eta >2.19$, $D= 18\pm2 m$, and $p_V = 0.59^{+0.25}_{-0.17}$. Fits with $\eta\sim\pi$ find a lower limit of $p_V\sim0.36$. The lack of a clear absorption band near 2.0 $\mu m$ shows that Kamo`oalewa is not a typical S- or V-type asteroid, but a faint feature (depth $3.0\pm0.5\%$) at $0.93\pm0.01$ $\mu m$ may indicate an affinity with high-Ca pyroxenes similar to E-type asteroids. If the shifted Band I center is indeed the result of a higher oldhamite abundance, the isotopic abundance of Ca within the mineral relative to the silicates can provide constraints on its origin \citep{Dai2024}. 

Measurements from Tianwen-2 regarding Kamo`oalewa's SED, elongated shape, and the obliquity of its rotation pole, will provide a ground-truth test of the asteroid's thermal properties and provide context for its extreme value of $\eta$. In-situ measurements of Kamo`oalewa's surface composition will provide direct tests of Kamo`oalewa's origins. Ground-truth observations will be key for validating remote interpretations of small NEAs generally.

Longer wavelength MIRI observations, as well as additional NIRSpec observations at multiple phase angles, could break thermal model degeneracies and provide stricter constraints on albdo and $\eta$. Along with the recent case study of 2024 YR4 \citep{Rivkin_2026}, JWST is ushering in a new era for rapidly characterizing the orbital and physical properties of NEAs that are similar in size to the Chelyabinsk impactor.

\begin{acknowledgments}
This work is based on observations made with the NASA/ESA/CSA James Webb Space Telescope. The specific observations analyzed can be accessed via doi:10.17909/dgwx-zz35. The data were obtained from the Mikulski Archive for Space Telescopes at the Space Telescope Science Institute, which is operated by the Association of Universities for Research in Astronomy, Inc., under NASA contract NAS 5-03127 for JWST. These observations are associated with program \#8663. Support for US investigators in program \#8663 was provided by NASA through a grant from the Space Telescope Science Institute, which is operated by the Association of Universities for Research in Astronomy, Inc., under NASA contract NAS5-03127. The authors wish to thank Jessica Sunshine, Bill Bottke, and Erik Asphaug for helpful discussions. We also thank an anonymous referee, whose reviews substantially improved this manuscript.
\end{acknowledgments}

\facilities{JWST, LBT}
\software{astropy \citep{2013A&A...558A..33A,2018AJ....156..123A,2022ApJ...935..167A}}  

\bibliography{sample701}{}
\bibliographystyle{aasjournalv7}

\begin{deluxetable*}{ccccc}\label{circumstances}
\digitalasset
\tablewidth{0pt}
\tablecaption{Per-dither observational circumstances and reflected irradiances \label{tab:description}}
\tablehead{
\colhead{Exposure Mid-Time} & \colhead{$\Delta$ (au)} & \colhead{r (au)} & \colhead{Phase ($^{\circ}$)} & \colhead{Integrated Irradiance (mJy)}
}
\startdata
2026-02-09 01:41:01.820 & 0.17340 & 1.06615 & 61.4763 & 1.766 $\pm$ 0.076 \\
2026-02-09 01:54:38.844 & 0.17340 & 1.06616 & 61.4722 & 1.803 $\pm$ 0.096 \\
2026-02-09 02:08:15.803 & 0.17340 & 1.06617 & 61.4681 & 1.965 $\pm$ 0.091 \\
2026-02-09 02:21:52.826 & 0.17339 & 1.06618 & 61.4639 & 1.879 $\pm$ 0.088 \\
2026-02-11 11:24:51.717 & 0.17256 & 1.06915 & 60.4358 & 1.133 $\pm$ 0.062 \\
2026-02-11 11:38:28.741 & 0.17256 & 1.06916 & 60.4317 & 1.493 $\pm$ 0.073 \\
2026-02-11 11:52:05.764 & 0.17256 & 1.06917 & 60.4277 & 1.228 $\pm$ 0.091 \\
2026-02-11 12:05:42.723 & 0.17255 & 1.06919 & 60.4237 & 1.616 $\pm$ 0.077 \\
2026-02-13 04:14:10.346 & 0.17195 & 1.07121 & 59.7166 & 2.133 $\pm$ 0.113 \\
2026-02-13 04:27:47.369 & 0.17195 & 1.07122 & 59.7127 & 1.830 $\pm$ 0.106 \\
2026-02-13 04:41:24.328 & 0.17195 & 1.07123 & 59.7087 & 2.102 $\pm$ 0.103 \\
2026-02-13 04:55:01.352 & 0.17194 & 1.07124 & 59.7047 & 1.876 $\pm$ 0.088 \\
2026-02-13 09:43:27.447 & 0.17187 & 1.07148 & 59.6211 & 1.519 $\pm$ 0.070 \\
2026-02-13 09:57:04.470 & 0.17187 & 1.07149 & 59.6171 & 1.234 $\pm$ 0.076 \\
2026-02-13 10:10:41.429 & 0.17186 & 1.07150 & 59.6132 & 1.430 $\pm$ 0.074 \\
2026-02-13 10:24:18.452 & 0.17186 & 1.07151 & 59.6092 & 1.225 $\pm$ 0.063 \\
2026-02-14 01:28:55.127 & 0.17163 & 1.07226 & 59.3483 & 1.541 $\pm$ 0.067 \\
2026-02-14 01:56:09.110 & 0.17162 & 1.07228 & 59.3405 & 2.436 $\pm$ 0.134 \\
2026-02-14 02:09:46.133 & 0.17162 & 1.07229 & 59.3366 & 1.803 $\pm$ 0.080 \\
2026-02-15 00:59:22.049 & 0.17127 & 1.07340 & 58.9458 & 1.152 $\pm$ 0.076 \\
2026-02-15 01:12:59.072 & 0.17127 & 1.07341 & 58.9420 & 1.456 $\pm$ 0.063 \\
2026-02-15 01:26:36.095 & 0.17127 & 1.07342 & 58.9381 & 1.356 $\pm$ 0.057 \\
2026-02-15 01:40:13.054 & 0.17126 & 1.07343 & 58.9343 & 1.604 $\pm$ 0.069 \\
2026-02-15 11:58:14.618 & 0.17111 & 1.07392 & 58.7598 & 1.753 $\pm$ 0.078 \\
2026-02-15 12:11:51.577 & 0.17110 & 1.07394 & 58.7559 & 1.806 $\pm$ 0.075 \\
2026-02-15 12:25:28.600 & 0.17110 & 1.07395 & 58.7521 & 1.708 $\pm$ 0.066 \\
2026-02-15 12:39:05.624 & 0.17110 & 1.07396 & 58.7483 & 1.211 $\pm$ 0.053 \\
\enddata
\tablecomments{Observational circumstances and brightness measurements. Integrated irradiance values were taken as the sum across each wavelength element between $0.7-3.5$ $\mu m$, which is dominated by reflected light.}
\end{deluxetable*}

\end{document}